\documentclass{article}
\usepackage{spconf,amsmath,graphicx}
\usepackage{dsfont}
\usepackage{amssymb}
\usepackage{subfigure}
\usepackage{wrapfig}
\usepackage{cite}
\usepackage{bm}
\usepackage{hyperref}
\usepackage{multirow}
\usepackage{tabularx}

\DeclareMathOperator*{\argmin}{arg\,min}

\title{Knowledge Transfer between Datasets for Learning-based Tissue Microstructure Estimation}
\name{ Yu Qin,  Yuxing Li,  Zhiwen Liu, Chuyang Ye$^{*}$ 
    	\thanks{${^{*}}$ Corresponding author: chuyang.ye@bit.edu.cn} 
		\thanks{{\bf{Acknowledgement.}} This work is supported by the National Natural Science Foundation of China (61601461), Beijing Natural Science Foundation (7192108), and Beijing Institute of Technology Research Fund Program for Young Scholars. The HCP-MGH dataset was provided by the MGH-USC Human Connectome Project and the Laboratory of Neuro Imaging at the University of Southern California. The HCP-Minn dataset was provided by the Human Connectome Project, WU-Minn Consortium and the McDonnell Center for Systems Neuroscience at Washington University.}
	}
\address{School of Information and Electronics, Beijing Institute of Technology, Beijing, China}

\begin{document}
%
\maketitle
\begin{abstract}
Learning-based approaches, especially those based on deep networks, have enabled high-quality estimation of tissue microstructure from low-quality \textit{diffusion magnetic resonance imaging} (dMRI) scans, which are acquired with a limited number of diffusion gradients and a relatively poor spatial resolution. 
These learning-based approaches to tissue microstructure estimation require acquisitions of training dMRI scans with high-quality diffusion signals, which are densely sampled in the $q$-space and have a high spatial resolution. 
However, the acquisition of training scans may not be available for all datasets.
Therefore, we explore knowledge transfer between different dMRI datasets so that learning-based tissue microstructure estimation can be applied for datasets where training scans are not acquired. 
Specifically, for a target dataset of interest, where only low-quality diffusion signals are acquired without training scans, we exploit the information in a source dMRI dataset acquired with high-quality diffusion signals. We interpolate the diffusion signals in the source dataset in the $q$-space using a dictionary-based signal representation, so that the interpolated signals match the acquisition scheme of the target dataset.
Then, the interpolated signals are used together with the high-quality tissue microstructure computed from the source dataset to train deep networks that perform tissue microstructure estimation for the target dataset. 
Experiments were performed on brain dMRI scans with low-quality diffusion signals, where the benefit of the proposed strategy is demonstrated.

\end{abstract}
\section{Introduction}
\label{sec:intro}

\textit{Diffusion magnetic resonance imaging}~(dMRI) noninvasively probes tissue microstructure, which provides valuable biomarkers for brain studies~\cite{johansen}.
However, typical dMRI scans are usually acquired with a limited number of diffusion gradients and a relatively poor spatial resolution. These low-quality diffusion signals could adversely affect the quality of tissue microstructure estimation for complex biophysical models using conventional model-based approaches~\cite{NODDI,AMICO}.

To improve tissue microstructure estimation, learning-based approaches have been developed to map low-quality diffusion signals to high-quality tissue microstructure~\cite{golkov,medn+,mesc,SR_q_DL}.
For example, in \cite{golkov} a \textit{multiple layer perceptron}~(MLP) is used to learn the mapping from diffusion signals acquired with a limited number of diffusion gradients to high-quality tissue microstructure. Since the diffusion signal associated with each diffusion gradient can be interpreted as a measurement in the $q$-space, this strategy is also termed as \textit{$q$-space deep learning}~($q$-DL). $q$-DL is then improved by \cite{medn+}, where a network is developed by exploiting the sparsity of diffusion signals in the $q$-space and it achieves improved results of tissue microstructure estimation. Spatial information can also be incorporated into the network design \cite{medn+,mesc}, where the estimation quality is further improved. These $q$-DL methods focus on the estimation of tissue microstructure with diffusion signals undersampled in the $q$-space. $q$-DL is then extended to \textit{super-resolved $q$-DL }~(SR-$q$-DL) in~\cite{SR_q_DL}, where the spatial resolution of tissue microstructure estimation results is improved as well. Specifically, the network design in~\cite{SR_q_DL} integrates the sparsity of diffusion signals in the $q$-space with super-resolution techniques, so that high-resolution tissue microstructure can be estimated from low-resolution dMRI scans acquired with a limited number of diffusion gradients.

In the learning-based approaches, the networks are trained by high-quality dMRI scans acquired with a large number of diffusion gradients (and a high spatial resolution for SR-$q$-DL), where high-quality training tissue microstructure can be computed with conventional model-based approaches.
However, in many situations, the datasets of interest may not contain high-quality training dMRI scans, and how to perform learning-based tissue microstructure estimation for such datasets is an open problem.

In this work, we explore knowledge transfer between datasets to allow learning-based tissue microstructure estimation for datasets where training scans are not acquired. We exploit datasets where high-quality dMRI scans are available. 
For convenience, such a dataset is referred to as the source dataset.
Then, we use the information in the source dataset to train a deep network that performs tissue microstructure estimation for the dataset of interest, which is referred to as the target dataset. 
However, in general the set of diffusion gradients of the target dataset is not a subset of that of the source dataset, and the diffusion signals in the source dataset cannot be directly used for network training.
Thus, to allow knowledge transfer from the source dataset to the target dataset, the diffusion signals of the source dataset are interpolated in the $q$-space so that the interpolated signals match the acquisition scheme of the target dataset and represent diffusion signals undersampled in the $q$-space in the target dataset.
The interpolation is performed with a dictionary-based representation of diffusion signals using the SHORE basis~\cite{merlet}.
The interpolated diffusion signals are used together with the high-quality training tissue microstructure computed from the original high-quality diffusion signals in the source dataset to train the estimation deep network for the target dataset.
The proposed strategy was evaluated on brain dMRI scans for both $q$-DL and SR-$q$-DL, where its benefit is shown.

\section{METHODS}
\label{sec:method}
Suppose we have a target dMRI dataset $\mathcal{D}_{\mathrm{t}}$, where the diffusion signals are acquired with a set $\mathcal{G}_{\mathrm{t}}$ of diffusion gradients that undersample the $q$-space and possibly a relatively low spatial resolution. 
We explore how to perform learning-based tissue microstructure for $\mathcal{D}_{\mathrm{t}}$ when training scans are not acquired for $\mathcal{D}_{\mathrm{t}}$.
We propose to address this issue by transferring knowledge from a source dataset $\mathcal{D}_{\mathrm{s}}$, where high-quality diffusion signals have been acquired, to the target dataset $\mathcal{D}_{\mathrm{t}}$. 
Specifically, in $\mathcal{D}_{\mathrm{s}}$, diffusion signals are acquired with a set $\tilde{\mathcal{G}}_{\mathrm{s}}$ of diffusion gradients densely sampling the $q$-space (and a high spatial resolution if high-resolution tissue microstructure estimation is desired). 
Using $\mathcal{D}_{\mathrm{s}}$, we train a deep network that matches the $q$-space sampling scheme of $\mathcal{D}_{\mathrm{t}}$, and tissue microstructure estimation can then be applied for $\mathcal{D}_{\mathrm{t}}$ with this deep network. 
The details of our strategy are described below.

\subsection{Signal interpolation in the $q$-space}
\label{ssec:interplation}
Since $\mathcal{G}_{\mathrm{t}}$ is generally not a subset of $\tilde{\mathcal{G}}_{\mathrm{s}}$, the diffusion signals in $\mathcal{D}_{\mathrm{s}}$ cannot be directly used to train the estimation network for $\mathcal{D}_{\mathrm{t}}$. Thus, we first interpolate the diffusion signals in $\mathcal{D}_{\mathrm{s}}$ in the $q$-space to generate undersampled diffusion signals that correspond to $\mathcal{D}_{\mathrm{t}}$. 
Note that for SR-$q$-DL, before the interpolation in the $q$-space, an additional step of creating low-resolution diffusion signals is required, so that the deep network can take low-resolution diffusion signals as input and output high-resolution tissue microstructure. The low-resolution diffusion signals are computed by taking the block-wise mean~\cite{tanno} in the scans in $\mathcal{D}_{\mathrm{s}}$. Specifically, if the upsampling rate in SR-$q$-DL is $\gamma$ (an integer), the size of the block for computing the low-resolution signals is $\gamma$.

We perform the signal interpolation in the $q$-space based on the fact that diffusion signals can be represented by a set of suitable basis functions~\cite{merlet}.
We select the SHORE basis~\cite{merlet}, which has been widely applied in the reconstruction of diffusion signals.
Specifically, suppose at a voxel the diffusion signal at the coordinate $\bm{q}$ in the $q$-space---i.e., associated with the diffusion gradient ${\bm{q}}$---is ${y(\bm{q})}$; then we have~\cite{merlet} 
\begin{eqnarray}
y({\bm{q}}) = \sum\limits_{n = 0}^N {\sum\limits_{l = 0}^L {\sum\limits_{m =  - l}^l {c_{nlm}}{{\Phi _{nlm}}({\bm{q}})} } } ,
\label{eq:1}
\end{eqnarray}
where $\Phi_{nlm}(\bm{q})$ is the value of the SHORE basis function with a radial order $n$, angular order $l$, and angular degree $m$ for the diffusion gradient $\bm{q}$, ${c_{nlm}}$ is the corresponding coefficient, and $N$ and $L$ represent the maximal radial order and angular order, respectively.
We can concatenate the diffusion signals in the $q$-space and rewrite Eq.~(\ref{eq:1}) in a matrix form for $\mathcal{D}_{\mathrm{s}}$ and $\mathcal{D}_{\mathrm{t}}$, which leads to the dictionary-based representation
\begin{eqnarray}
{\bm{y}_{\mathrm{s}}} = {{\bm{\Phi }}_{\mathrm{s}}}{{\bm{c}}_{\mathrm{s}}},
{\bm{y}_{\mathrm{t}}} = {{\bm{\Phi }}_{\mathrm{t}}}{{\bm{c}}_{\mathrm{t}}}.
\label{eq:2}
\end{eqnarray}
Here, ${\bm{y}_{\mathrm{s}}}$ and ${\bm{y}_{\mathrm{t}}}$ represent the diffusion signal vector at a voxel in $\mathcal{D}_{\mathrm{s}}$ and $\mathcal{D}_{\mathrm{t}}$, respectively, ${{\bm{\Phi }}_{\mathrm{s}}}$ and ${{\bm{\Phi }}_{\mathrm{t}}}$ are the dictionaries computed from the SHORE basis functions corresponding to $\tilde{\mathcal{G}}_{\mathrm{s}}$ and $\mathcal{G}_{\mathrm{t}}$, respectively, and ${{\bm{c}}_{\mathrm{s}}}$ and ${{\bm{c}}_{\mathrm{t}}}$ are the corresponding vectors of coefficients.

Using Eq.~(\ref{eq:2}), we can interpolate the signals in the $q$-space for $\mathcal{D}_{\mathrm{s}}$, so that the interpolated signals correspond to the $q$-space sampling scheme in $\mathcal{D}_{\mathrm{t}}$. To achieve that, we first estimate ${{\bm{c}}_{\mathrm{s}}}$ by solving the following minimization problem~\cite{merlet} 
\begin{eqnarray}
\argmin\limits_{{\bm{c}}_{\mathrm{s}}} ||{\bm{y}_{\mathrm{s}}} - {{\bm{\Phi }}_{\mathrm{s}}}{{\bm{c}}_{\mathrm{s}}}|{|_{{\ell_{2}}}} + {\lambda _l}||{\mathbf{L} {{\bm{c}}_{\mathrm{s}}}}|{|_{{\ell_{2}}}} + {\lambda _n}||{\mathbf{N} {{\bm{c}}_{\mathrm{s}}}}|{|_{{\ell_{2}}}},
\label{eq:3}
\end{eqnarray}
where ${\lambda _l}$ and ${\lambda _n}$ are weighting constants, and ${\mathbf{N}}$ and ${\mathbf{L}}$ are two diagonal matrices such that $diag({\mathbf{N}}) = n(n + 1)$ and $diag({\mathbf{L}}) = l(l + 1)$~\cite{merlet}.
Eq.~(\ref{eq:3}) has a closed form solution~\cite{merlet}
\begin{eqnarray}
\hat{\bm{c}}_{\mathrm{s}} = {({\bf{\Phi }_{\mathrm{s}}^{\mathsf{T}}} {\bf{\Phi }_{\mathrm{s}}} + {\lambda _l}{{\mathbf{L}}^{\mathsf{T}}}{\mathbf{L}} + {\lambda _n}{{\mathbf{N}}^{\mathsf{T}}}{\mathbf{N}})^{ - 1}}{{\bm{\Phi }_{\mathrm{s}}^{\mathsf{T}}} \bm{y}_{\mathrm{s}}}.
\label{eq:4}
\end{eqnarray}
The default parameters given by Dipy~\cite{dipy} are used for the SHORE representation and solving Eq.~(\ref{eq:3}).
Then, with the estimated coefficients $\hat{\bm{c}}_{\mathrm{s}}$, we compute the interpolated diffusion signals corresponding to $\mathcal{G}_{\mathrm{t}}$ by $\hat{\bm{y}}_{\mathrm{s}} = \mathbf{\Phi}_{\mathrm{t}}\hat{\bm{c}}_{\mathrm{s}}$.

\subsection{Network training and evaluation}
\label{sssec:train}
Using the interpolated signals and high-quality tissue microstructure computed from the original high-quality diffusion signals in $\mathcal{D}_{\mathrm{s}}$, we can train a deep network for tissue microstructure estimation for $\mathcal{D}_{\mathrm{t}}$. We consider patch-based deep networks for tissue microstructure estimation, which have achieved state-of-the-art performance. In particular, we select the networks in~\cite{medn+} and~\cite{SR_q_DL} for $q$-DL and SR-$q$-DL, respectively. These deep networks take patches of low-quality diffusion signals as input and output high-quality tissue microstructure at the center voxel(s) of the input patch. The outputs are concatenated to obtain the final tissue microstructure map. We train the deep networks according to their proposed strategies as described below.

From $\mathcal{D}_{\mathrm{s}}$ we can extract training samples by using patches of which the centers are inside the brain. For convenience, we denote the set of training samples by $\mathcal{S}{{ = \{ }}\hat{\bm{y}}_{\mathrm{s}}^{(i)}{, \bm{m}}_{\mathrm{s}}^{(i)}{\rm{\} }}_{i = 1}^M$, where $M$ is the total number of training samples, and $\hat {\bm{y}}_{\mathrm{s}}^{(i)}$ and ${\bm{m}}_{\mathrm{s}}^{(i)}$ are the interpolated signals in the input patch and high-quality tissue microstructure of the output patch for the $i$-th training sample, respectively. For $q$-DL, we follow~\cite{medn+} and set the patch sizes of $\hat {\bm{y}}_{\mathrm{s}}^{(i)}$ and ${\bm{m}}_{\mathrm{s}}^{(i)}$ to $3^3$ and $1^3$, respectively; for SR-$q$-DL, we follow~\cite{SR_q_DL} and set the patch sizes of $\hat {\bm{y}}_{\mathrm{s}}^{(i)}$ and ${\bm{m}}_{\mathrm{s}}^{(i)}$ to $5^3$ and $2^3$, respectively.
Note that in SR-$q$-DL, the diffusion signals are downsampled in the spatial domain as describe in Sect.~\ref{ssec:interplation} before $q$-space interpolation, and training tissue microstructure is computed at its original resolution.
We follow the settings of training in~\cite{medn+} and~\cite{SR_q_DL} for $q$-DL and SR-$q$-DL, respectively, and the deep networks trained by $\mathcal{S}$ can then be used to estimate tissue microstructure for $\mathcal{D}_{\mathrm{t}}$.


For evaluation, like in~\cite{medn+} and~\cite{SR_q_DL}, we considered the tissue microstructure measures described by the NODDI model~\cite{NODDI}, which has been widely used in neuroscientific studies. These measures include the intra-cellular volume fraction~$v_{\mathrm{ic}}$, \textit{cerebrospinal fluid}~(CSF) volume fraction~$v_\mathrm{iso}$, and \textit{orientation dispersion}~(OD). The AMICO algorithm~\cite{AMICO} was used to efficiently compute the training NODDI tissue microstructure using the original high-quality diffusion signals in $\mathcal{D}_{\mathrm{s}}$.
For quantitative evaluation, high-resolution diffusion signals densely sampling the $q$-space were acquired for $\mathcal{D}_{\mathrm{t}}$, and from these signals gold standard tissue microstructure (used for evaluation only) was computed using AMICO~\cite{AMICO}.
The absolute difference between the estimation result and the gold standard was computed to measure the estimation error. 

\section{RESULTS}
\label{sec:results}

We selected the HCP-MGH dataset~\cite{HCP_MGH} as the target dataset, where dMRI scans of 32 subjects were used for evaluation. The dMRI scans were acquired with an isotropic spatial resolution of 1.5 mm and 512 diffusion gradients ($b=1000,3000,5000,10000~\mathrm{s}/\mathrm{mm}^2$). To generate low-quality dMRI scans with diffusion signals undersampled in the $q$-space, 36 diffusion gradients (18 on each of the shells $b=1000,3000~\mathrm{s}/\mathrm{mm}^2$) were selected as the diffusion gradients $\mathcal{G}_{\mathrm{t}}$ that undersample the $q$-space. 
For $q$-DL, we sought to perform high-quality tissue microstructure estimation using these undersampled signals corresponding to $\mathcal{G}_{\mathrm{t}}$.
For SR-$q$-DL, these undersampled diffusion signals were then subsampled in the spatial domain by a factor $\gamma$ = 2 using the block-wise mean~\cite{tanno}, so that low-resolution dMRI scans with a limited number of diffusion gradients were generated; and we sought to estimate high-resolution (an upsampling rate $\gamma$ = 2) tissue microstructure from these low-resolution scans. 
For both $q$-DL and SR-$q$-DL, the original high-resolution diffusion signals densely sampled in the $q$-space were only used to compute the gold standard tissue microstructure for evaluation and not used in network training. 

We selected the HCP-Minn dataset~\cite{HCP_Minn} as the source dataset, where dMRI scans of five subjects were used for network training. The dMRI scans were acquired with an isotropic spatial resolution of 1.25 mm and 270 diffusion gradients ($b=1000,2000,3000 ~\mathrm{s}/\mathrm{mm}^2$). 
Using these scans, we computed high-quality tissue microstructure maps for training. 
Then, we generated low-quality input diffusion signals for network training.
For SR-$q$-DL, the dMRI scans were subsampled in the spatial domain by a factor $\gamma$ = 2 using the block-wise mean~\cite{tanno}. Then, $q$-space interpolation was performed for the dMRI scans with or without spatial subsampling for SR-$q$-DL or $q$-DL, respectively. Finally, deep networks were trained for the target dataset using training samples extracted from the source dataset.
Note that to make the gold standard of the target dataset comparable with the training tissue microstructure in terms of the number of diffusion gradients and $b$-value range, the diffusion signals associated with $b=10000 ~\mathrm{s}/\mathrm{mm}^2$ in the target dataset were not used for the computation of gold standard for evaluation. 

\begin{figure}[!t]
	\centering
	\includegraphics[width=1.0\columnwidth]{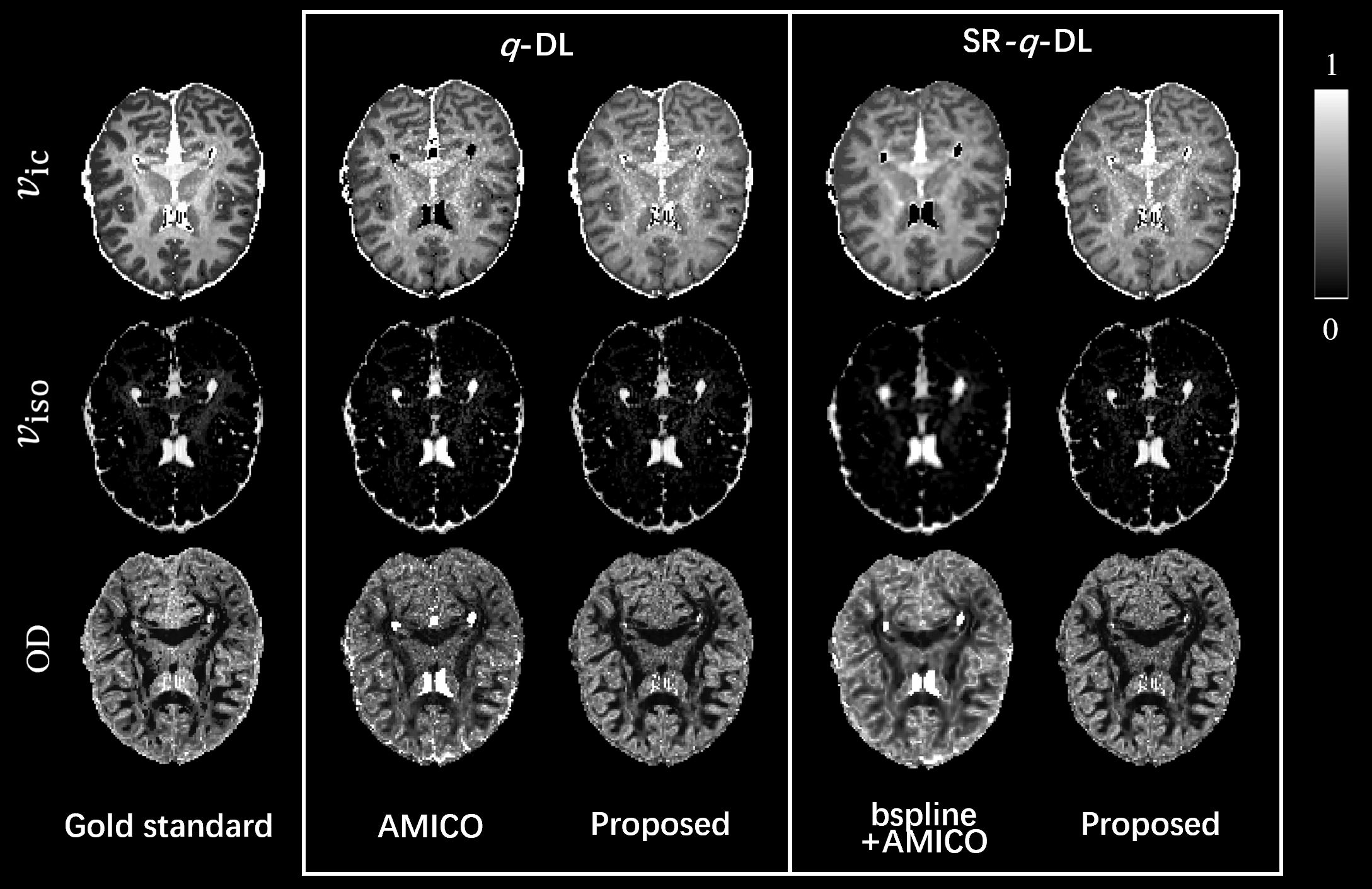}
	\caption{A representative axial view of the tissue microstructure maps estimated by the proposed method. The gold standard and the AMICO results are also shown for reference.}
	\label{fig:results}
\end{figure}

\begin{figure}[!t]
	\centering
	\includegraphics[width=1.0\columnwidth]{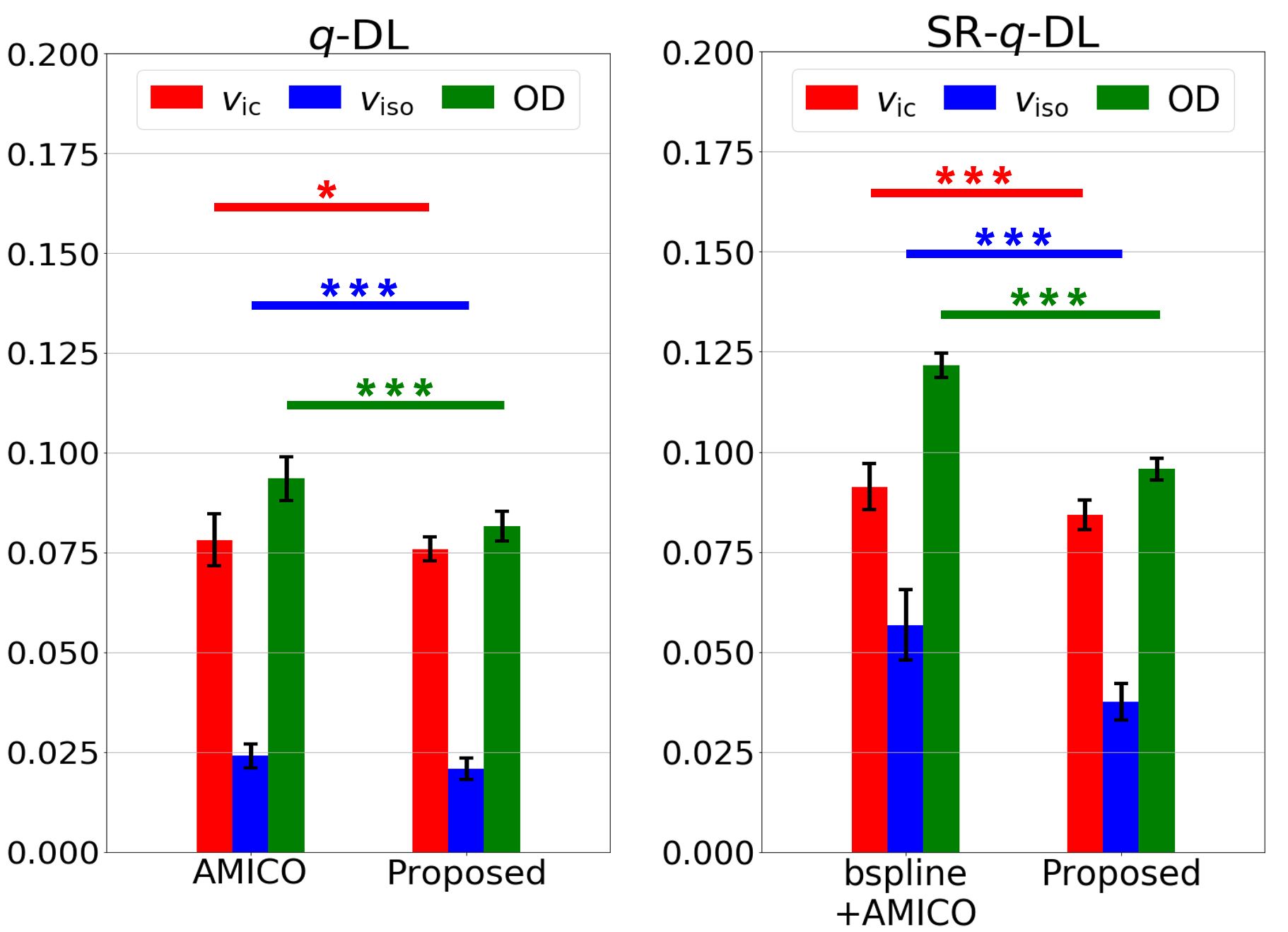}
	\caption{Means and standard deviations of the average estimation errors in the brains of test subjects for $q$-DL and SR-$q$-DL. Asterisks indicate that the difference between the proposed method and the competing method is significant using a paired Student's $t$-test. ($^{***}p<0.001$, $^{*}p<0.05$)}
	\label{fig:stat}
\end{figure}

We first qualitatively show that the proposed strategy produces sensible high-quality tissue microstructure maps. The estimation results of $q$-DL and SR-$q$-DL on a representative test subject are shown in Fig.~\ref{fig:results} together with the gold standard. The estimation results are also compared with the conventional method which does not require acquisitions of training data for the target dataset. 
Specifically, for $q$-DL we consider the AMICO results computed from the diffusion signals undersampled in the $q$-space, and for SR-$q$-DL we consider the AMICO results estimated by upsampling the low-quality dMRI scans by a factor of two using 3rd-order bsplines. 
Our method produced high-quality tissue microstructure maps that restored more anatomical details than the conventional approach.
Next, the estimation results were evaluated quantitatively.  The average estimation errors in the brain (excluding CSF~\cite{medn+}) of each test subject were computed, and the means and standard deviations of the average errors are summarized in Fig.~\ref{fig:stat}. The proposed method was also compared with AMICO using paired Student’s $t$-tests, and the errors of our method are significantly smaller than those of AMICO, as indicated by the asterisks in Fig.~\ref{fig:stat}.

\section{CONCLUSION}
\label{sec:conclusion}

We have explored knowledge transfer between datasets for performing learning-based tissue microstructure estimation for datasets where training scans are not acquired. Using a dictionary-based signal representation, we interpolate the high-quality diffusion signals in the $q$-space for the source dataset to match the acquisition scheme of the target dataset. The interpolated signals allow the training of deep networks that estimate high-quality tissue microstructure for the target dataset. The proposed method was applied to $q$-DL and SR-$q$-DL, and it was compared with conventional model-based estimation approaches. Results on brain dMRI scans demonstrate the benefit of the proposed strategy.




\bibliographystyle{IEEEbib}
\bibliography{strings,refs}

\end{document}